\documentstyle[preprint,eqsecnum,aps,prb]{revtex}

\begin{document}
\title{Interfacial strain measurements in $SrRuO_3/SrMnO_3$ magnetic multilayers.}
\author{P. Padhan, W. Prellier\thanks{%
prellier@ismra.fr} and B.\ Mercey}
\address{Laboratoire CRISMAT, CNRS\ UMR 6508, ENSICAEN, 6 Bd du Mar\'{e}chal Juin,\\
F-14050 Caen Cedex, FRANCE.}
\date{\today}
\maketitle

\begin{abstract}
Magnetic multilayers of $(SrRuO_3)_m(SrMnO_3)_n$ were grown artificially
using the pulsed laser deposition technique on ($001$)-oriented $SrTiO_3$
substrates. The state of strain at the interfaces and the structural
coherency are studied in details utilizing asymmetrical $X$-ray diffraction
and the $\sin ^2\psi $ method. First, the evolution of the lattice
parameters, the crystallinity and the epitaxy of the films are evaluated as
a function of the number of $SrMnO_3$ unit cells using $X$-rays diffraction
and transmission electron microscopy.\ Second, our results on the stress
indicate that the $SrRuO_3/SrMnO_3$ superlattices show a larger residual
strain as compared to the single layer film of $SrRuO_3$. This suggests that
the lattice stiffening from interfacial strain and inhibiting the
dislocation by composition modulation. Finally, these results bring insights
on the interfacial stress measurements of oxide multilayers that can be used
to control the physical properties at the level of the atomic scale.
\end{abstract}

\newpage

Magnetic multilayer structures based on transition metals \cite{1,2} and
their compounds \cite{1,2,3,4,5,6,7,8,9,10,11,12,13,14} have high potential
for technological applications as their transport and magnetic properties
can be controlled with the non-magnetic spacer layer thickness. However, to
use these materials for applications, it is necessary to understand and
control precisely the physical properties that depend on various parameters
such as the layer materials, their thicknesses and the interfaces between
them. In the case of magnetic multilayers, the interfaces are rich in
magnetic and structural coordinations. Moreover, the lattice mismatch and
thickness between the two constituent materials will also modify the
strength of the interfaces. Furthermore, the lattice mismatch induced-strain
changes the physical properties of the oxide thin films, including the
transition temperature in high-temperature superconductors \cite{15,15a} and
in ferroelectric oxides \cite{16}. Similar effect in the Mn-based
multilayers is responsible for significant variation in magnetization as
well as in electronic, transport and structural properties \cite{13,14,16a}.
For example, Kreisel {\it et al.} \cite{17} have observed tensile-strain
induced rhombohedral-to-orthorhombic phase transition in $%
La_{0.7}Sr_{0.3}MnO_3/SrTiO_3$ by Raman scattering. In this system, these
two phases $La_{0.7}Sr_{0.3}MnO_3$ and $SrTiO_3$ coexist in the superlattice
with intermediate range of layer thickness. Y. Lue and coworkers \cite{10}
have also studied structural and transport properties of $%
La_{2/3}Ba_{1/3}MnO_3/SrTiO_3$ structure. They observed that electrical
transport properties of these samples strongly depend on the strain-induced
distortion in the $La_{2/3}Ba_{1/3}MnO_3$ layer.

Considering the above points, it is interesting first to fabricate magnetic
multilayers using the thin film deposition processes. Second, the artificial
control of their properties as a function of the spacer layer thickness is
required. Third, the interfacial stress that plays an important role upon
the structural and magneto-transport properties needs to be evaluated.

In this article, we report the structural study of the superlattices
consisting of $20$ $unit$ $cells$ ($u.c.$) thick $SrRuO_3$ ($SRO$) and $n$ $%
u.c.$ thick $SrMnO_3$ ($SMO$) where $n$ varies from $1$ to $20$ grown on ($%
001$)-oriented $SrTiO_3$ ($STO$, cubic with $a=3.905$ $\AA $). We choose
these materials because $SRO$\ is a ferromagnetic metal\cite{18} whereas $%
SMO $ is a highly insulating antiferromagnet\cite{SMO}. Moreover, the
lattice parameter of bulk $SRO$ ($a_{SRO}$ $=$ $3.93$ $\,\AA $) is larger
than $a_{STO}$ with a lattice mismatch $+$ $0.6$ $\%$ whereas the lattice
parameter of $SMO$ ($a_{SMO}$ $=$ $3.805$ $\,\AA $) is smaller than $a_{SRO}$
with lattice mismatch $-$ $3.0$ $\%$. Though there is a large lattice
mismatch between $SRO$ and $SMO,$ we have chosen this combination because
the $A$-site ions are the same and the reduction of $B$-site distortion \cite
{17a} is expected at the interfaces between $SRO$\ and $SMO$. The state of
strain at the interfaces and the structural coherency are studied using the $%
\sin ^2\psi $ method, and our results are reported in this article. The
superlattices show larger residual strain compared to the single layer film
of $SRO$, suggesting that the lattice stiffening from interfacial strain and
inhibiting dislocation by composition modulation.

A multitarget pulsed laser deposition system\cite{15a} was used to grow $SRO$
thin films and $SRO/SMO$ superlattices on ($001$)- $SrTiO_3$ substrates. The
thin films of $SRO$ and the superlattices were deposited at $720$ $^{\circ }C
$ in oxygen ambient of $30$ $mtorr$. The deposition rates (typically $%
\symbol{126}0.26$ $\AA /pulse$) of $SRO$ and $SMO$ were calibrated for each
laser pulse of energy density $\symbol{126}3$ $J/cm^2$. After the deposition
the chamber was filled to $300$ $torr$ of oxygen at a constant rate, and
then the samples were slowly cool down to room temperature at the rate of $20
$ $^{\circ }C/\min $. The superlattice structures were synthesized by
repeating $15$ times the bilayer comprising of $20$ $u.c.$ $SRO$ and $n$ $%
u.c.$ $SMO$. In all samples $SRO$ is the bottom layer, and the modulation
structure was covered with $20$ $u.c.$ $SRO$ to keep the structure of the
top $SMO$ layer stable. These periodic modulation in composition was created
on the basis of established deposition rates of $SRO$ and $SMO$ were
confirmed from the positions of superlattice reflections in $X$-ray $\theta
-2\theta $ scans. The epitaxial growth and the structural characterization
of the multilayers and single layer films were performed using $X$-ray
diffraction, electron dispersive spectroscopy ($EDS$) and transmission
electron microscopy ($TEM$). The $\theta -2\theta $, $\Phi $ and $\omega $%
-scans were performed using $Seifert$ $XRD$ $3000P$ and $Philips$ $MRD$ $%
X^{\prime }pert$ diffractometers ($\lambda $ $=$ $1.54069$ $\AA $). The $TEM$
is a $JEOL$\ $2010$ with a point resolution of $1.8$ $\AA $. Resistivity ($%
\rho $) was measured as a function of temperature ($T$) in PPMS\ Quantum
Design.

In bulk form $SRO$ exhibits only pseudocubic perovskite structure\cite{18}.
In contrast, stoichiometric $SMO$ crystallizes in cubic as well as hexagonal
phase\cite{19}. The cubic perovskite structure of $SMO$ is not stabilized in
its single layer thin film form; however, our results of $X$-ray diffraction
and transmission electron microscopy show the formation of cubic perovskite
structure of $SMO$ layer in the superlattices as previously observed\cite
{16a,19}. This result indicates that, $SMO$\ can be stabilized as a cubic
structure between two $SRO$\ layers\cite{19a}.

Our samples with alternate layers of $SRO$ and $SMO$ on $STO$ show ($00l$)
diffraction peaks of the constituents, indicates the growth of epitaxial
pseudocubic phase with the $c$-axis orientation, {\it i.e.}, $c$-axis
perpendicular to the substrate plane. In Fig.1, we show the $\theta -2\theta 
$ scan for several samples with different spacer layer thickness. These
scans are around the ($002$) reflection ($42{{}^{\circ }}$ $-$ $49{{}^{\circ
}}$ in $2\theta $) of these pseudocubic perovskites. As the $SMO$ layer
thickness increases above $1$ $u.c.$, the fundamental ($002$) diffraction
peak of the constituents shifted towards the angular position of the $STO$
and overlap it for $n$ $>$ $10$. The sample with $n$ $=$ $1$ shows two weak
satellite peaks on the lower angle side of the ($002$) diffraction peak of
the constituents. The presence of higher order strong satellite peaks on
either side of the ($002$) diffraction peak for samples with $n$ $\geq $ $2$
clearly indicate the formation of a new structure having a periodic chemical
modulation of the constituents.

In $SRO/SMO$ superlattices, the two constituents have perovskite structure
and the difference in the lattice parameters between them is significant ($%
3.93\AA $ vs. $3.805\AA $). Also the atomic scattering factor of $Ru$ is
higher than $Mn$. The higher order satellite peaks with strong intensity is
expected to be observed in the $X$-ray diffraction. To extract the
information about the coherency at the interfaces and the periodic chemical
modulation ($\Lambda $) of these superlattices from $\theta -2\theta $
scans, we have carried out quantitative refinement of the superlattice
structure using $DIFFaX$ program\cite{20}. The experimental and simulated
diffraction profiles of the sample with $n$ $=$ $5$ is shown in Fig. 2(a).
It shows only the $2\theta $-range close to the fundamental ($002$)
reflection ($42{{}^{\circ }}$ $-$ $51{{}^{\circ }}$ in $2\theta $). The
simulated profile is in good agreement with the measured $\theta -2\theta $
scan with respect to the satellite peak position and relative intensity
ratio. The inset in Fig.2(a) shows the rocking curve ($\omega $-scan)
recorded around the fundamental ($002$) diffraction peak of the sample with $%
n$ $=$ $5$. The full-width-at-half-maximum ($FWHM$) of the rocking curve is $%
0.125{{}^{\circ }}$, close to the instrumental limit, suggesting a high
crystalline quality of the structure in the samples. The $FWHM$ of the
rocking curve also correlates the structural coherence length $\xi $ of the
sample with the relation $\xi $ $=$ $\frac{2\pi }{Q\text{ }.\text{ }FWHM}$ 
\cite{21}, where $Q$ ($\approx $ $\frac 1d$) is the scattering vector length
and $FWHM$ is in radians. The coherence length of the sample in the
out-of-plane direction is nearly the same as the total thickness of the
multilayer structure, confirming the coherency and the single crystallinity
of the samples.

An asymmetric diffractometer configuration provides quantitative measure of
the in-plane coherency, pseudomorphic growth and the stress in all three
directions. In this configuration, the $\Phi $-scan of the sample with $n$ $%
= $ $5$ from the asymmetric \{$103$\} planes is shown in the Fig. 2(b). The $%
\Phi $-scans of the substrate and film correspond to the angular position of
the substrate and the constituents in the $\theta -2\theta $ scan at
asymmetric \{$103$\} planes. The presence of symmetric and periodic peaks
with a period of $90^{\circ }$ confirms the four-fold symmetry of these
pseudocubic perovskites. The negligibly small difference between the angular
position of the peak (in the $\Phi -$scan) of the substrate and the film
clearly shows the cube-on-cube growth morphology of the film.\ The in-plane
alignment is as follows: $[100]_{STO}//[100]_F$ and $[010]_{STO}//[010]_F$
(where the index F refers to the film).

The quality of the superlattices is confirmed by the electron diffraction ($%
ED$) study.\ An example of an $ED$ cross section, for a ($SRO$)$_{20}$($SMO$)%
$_5$ superlattice, is given in Fig.3(a). Note that the $ED$ is a
superposition of $SRO$\ and $SMO$ . The perfect $ED$ patterns confirms the $%
c $-axis orientation of the superlattice and, also, the perovskite
structure. Moreover, the satellite spots (see inset of Fig.3(a)), due to the
periodic stacking of the $SRO$\ and $SMO$ layers, are clearly visible.\ The
corresponding cross-section high resolution electron microscopy ($HREM$)\
image is shown in Fig.3(b). It confirms the presence of superstructure and
sharp heteroepitaxial $SRO$ $-$ $SMO$\ interfaces. The image also indicates
that the $SMO$ perovskite-type is stabilized between two $SRO$\ layers, and
actually, adopts a pseudocubic structure \cite{19a}.

Having the epitaxial and pseudocubic growth morphology, it is necessary to
verify the periodicity of all samples with different spacer layer thickness.
In ($20$ $u.c.$) $SRO/$($n$ $u.c.$) $SMO$ structure, the average
superlattice period is:

\begin{description}
\item  1. \qquad $\frac \Lambda {20+n}=$ $\left( \frac{20\times a_{SRO}\text{
}+\text{ }n\times a_{SMO}}{20\text{ }+\text{ }n}\right) =$ $\left( \frac{%
20\times a_{SRO}\text{ }+\text{ }n\times a_{SMO}}N\right) $.
\end{description}

Where $N$ $=$ $20$ $+$ $n$. Since the fundamental diffraction peak of the
superlattice is due to the diffraction from the constituent, we have assumed
the lattice parameter of the superlattice ($a$) as $a$ $\simeq \frac{a_{SRO}%
\text{ }+\text{ }a_{SMO}}2$. The superlattice period can be expressed as:

\begin{description}
\item  2. $\qquad \Lambda $ $\simeq $ ($20$ $+$ $n$) $\times $ $a$.
\end{description}

For higher spacer layer thickness (i.e. $n$ $=$ $20$) superlattice period is 
$\simeq $ $156$ $\AA $. This suggests that the coherence length of the
sample is much higher than the superlattice period. Therefore, the higher
angle satellite peak positions\cite{22} can be indexed about `$a$'; $\frac{%
2\sin \theta }\lambda $ $=$ $\frac 1a$ $\pm $ $\frac n\Lambda $, where $%
\theta $ is the angular position of the satellite peak and $\lambda $ is the 
$X$-ray wavelength. We used the following equation \cite{22} to extract the
superlattice period from the satellite peak positions in the $\theta
-2\theta $ scan:

\begin{description}
\item  3. \qquad $\Lambda $ $=$ $\frac \lambda {2\text{ }\times \left( \sin
\theta _i\text{ }-\text{ }\sin \theta _{i+1}\right) }$,
\end{description}

where $\theta _i$ and $\theta _{i+1}$ are the angular position of the $ith$
and ($i+1$)$th$ order satellite peak, respectively. The calculated values of 
$\Lambda $ from the different successive satellite peak positions is given
in Fig. 4(a), for different values of $n$. The superlattice period is linear
with $n$ and follows eq. (2), indicating a high quality of the different
samples and a clear correlation as a function of the spacer layer thickness.

As previously stated, the physical properties of magnetic thin films ($Mn$%
-based system) are strongly dependent on the strains imposed by the substrate%
\cite{22a}. This dependence has also been reported in the case of $SRO$ thin
films\cite{23} where $a_{SRO}$ is larger than $a_{STO}$ which indicates the
presence of compressive in-plane stress on the $SRO$ film. The
substrate-induced stress modifies the interatomic distance in $SRO$ and this
is maximum close to the $STO$\ substrate. However, the substrate-induced
stress relaxed as the number of $SRO$ layers increases. This is evidenced
when the lattice parameter of $SRO$ approaches to bulk value. Since $a_{SMO}$
is smaller than the $a_{SRO}$, the $SMO$ layer on $SRO$ will experience a
tensile strain within the plane. Consequently, the strain of the $SRO/SMO$
superlattices is a combine effect of substrate-induced strain as well as the
strain originated from the interfaces.

In this superlattice system, these strains are opposite in nature
(substrate-induced strain is compressive, whereas the strain at the
interfaces is tensile). Also the lattice parameters calculation suggest that
the interfacial strain is larger compare to the substrate induced strain.
So, it is important to understand both the influence of strain on the
lattice parameter of this structure and the influence of the $SMO$ layer
thickness upon the strain. In Fig.4(b), we report the average out-of-plane
lattice parameter of various samples as a function of spacer layer
thickness. The out-of-plane lattice parameter of $20$ $u.c.$ thick $SRO$ on $%
STO$ is $4.05\AA $, while it is $4.003\AA $ for the superlattice with $1$ $%
u.c.$ spacer layer. From the figure, we observed that as the spacer layer
thickness increases, the out-of-plane lattice parameter of the superlattice
decreases and approaches the bulk value of $SRO$ indicate a smooth
relaxation of the strain within the film.

In the transition metal multilayers each layer of the constituent has single
element where the lattice mismatch leads to the planar deformation at the
interfaces and hence its structure. While in a multilayer designed from
various transition metal compounds, the lattice mismatch introduces a
deformation in the $3D$-coordination of the transition metal element. To
understand the structural correlation of this $SRO/SMO$ system at the
interfaces, we have studied the asymmetric reflection of these samples using
the conventional $\sin ^2\psi $ method \cite{24} (where $\psi $ is the angle
between the lattice plane normal and the sample surface normal). This method
is commonly used to calculate Poisson's ratio ($\nu $), in-plane and
out-of-plane strain, and the strain free lattice parameter of the films. The
lattice mismatch between the deposited material and the substrate is the
source of strain in epitaxial thin film. In addition, the strain ($%
\varepsilon $) of the film along the direction of diffraction [$hlk$] from
any $hkl$ reflection for a biaxial strain state is defined as \cite{24}:

\begin{description}
\item  4. $\varepsilon $ $=$ $\frac{d_{hkl}\text{(}\phi \psi \text{) }-\text{
}d_o\text{ }}{d_o}$ $=$ ($\varepsilon _{11}$ $-$ $\varepsilon _{33}$) $\sin
^2\psi $ $+$ $\varepsilon _{33}$,
\end{description}

where $\phi $ is the angle between the projected lattice plane normal and an
in-plane axis. The parameters $d_{hkl}$($\phi \psi $) and $d_o$ are the
strained and un-strained (i.e. bulk value) ($hkl$) plane spacing of the
sample, respectively. $\varepsilon _{11}$ $=$ $\varepsilon _{22}$ are the
in-plane strain components and $\varepsilon _{33}$ is the out-of-plane
strain component in the film. The values of $d_o$ and $\varepsilon _{ii}$ ($i
$ $=$ $1$, $2$, $3$) depend on the elastic constant (or Young's modulus, $E$%
) and $\nu $.

We have chosen a unique direction with constant $h$ and $k$ to measure the
diffracted $X$-ray intensity as well as $\psi $ from $10l$ ($l$ $=$ $1$, $2$%
, $3$ and $4$) asymmetric reflection. The value of $\psi $ is sensitive to
the alignment of the sample, and to avoid the misaligned contribution of $%
\psi $, we have averaged over all $\phi $-directions. In Fig.5(a) we show
the $d_{10l}$($\phi \psi $) vs. $\sin ^2\psi _{10l}$ plot for two samples ($%
n=1$ and $n=12$). The values are similar for both the samples whose strain
free lattice parameter of the bilayer is expected to be different. From the
experimental view point the values of $\psi $ is also expected to be the
same value for a known plane in each sample.

Assuming the same strain free lattice parameter for all samples, we have
calculated the in-plane and out-of-plane strain from the $d_{10l}$($\phi
\psi $) vs. $\sin ^2\psi _{10l}$ plot. We have determined the value of $d_o$
from the Fig. 5(a) at $\sin ^2\psi _0$ $=$ $\frac{2\nu }{1+\nu }$, using the 
$\nu $ value ($\nu =0.327$ in agreement with previous reports on manganite
thin films\cite{24a}) calculated from the [$111$] direction. The value of $%
\nu $ was calculated using the relation\cite{24}:

\begin{description}
\item  5. ~ $(\frac{a_F-a_{STO}}{a_F})=(\frac{c_F-c_{STO}}{c_F})\times (%
\frac{1-\nu }{1+\nu })$
\end{description}

where $a_F$ and $c_F$ are the $a$-axis and $c$-axis lattice parameters of
the film ($a_{STO}=c_{STO}=3.905\AA $). These values $a_F$ and $c_F$ are
calculated from the [$111$] direction for the multilayer with $n=1$. Using
this value of $d_o$ in the eq. 4, we have calculated the strain components
for different samples are shown in Fig.5(b). The $\varepsilon _{11}$ and $%
\varepsilon _{33}$ are opposite in nature and the in-plane strain is
stronger as expected from lattice parameter consideration. From this figure,
we found that the strain is independent of the superlattice period although
the out-of-plane lattice parameter shows the relaxation of the stress at the
higher spacer layer thickness (see Fig.4(b)). However, this analysis does
not distinct the strain in the multilayer and the single layer $SRO$ film.
Also the ($111$) diffraction peak of the sample overlaps with that of the
substrate peak which prohibits calculating $d_o$ of each sample.

The eq. 4, is applicable for a thin film where the structure has a single
interface between the film and the substrate. In the case of multilayer,
which has more than one interfaces, its lattice parameter depends on the
thickness of the bilayer. We have assumed $d_{hkl}$($\phi \psi $) $=$ $a_f/%
\sqrt{h^2\text{ }+\text{ }k^2\text{ }+\text{ }l^2}$, where $a_f$ is the
average lattice parameter of the bilayer. The average lattice parameter of
the bilayer can be expressed as a function of $N$:

\begin{description}
\item  6a. $a_f$ $=$ $a_{SMO}$ + $\frac{20}N\times $($a_{SRO}$ $-$ $a_{SMO}$)
\end{description}

Using this value of $a_f$, eq. 4 can be written as:

\begin{description}
\item  6b. $\frac{[a_{SMO}\text{ }+\text{ }(\frac{20}N)\text{ }(a_{SRO}\text{
}-\text{ }a_{SMO})]}{\sqrt{h^2\text{ }+\text{ }k^2\text{ }+\text{ }l^2}}$ $=$
$d_o$($\varepsilon _{11}$ $-$ $\varepsilon _{33}$) $\sin ^2\psi $ $+$ $d_o$ (%
$\varepsilon _{33}$ $+$ $1$)
\end{description}

To apply this relation to the $SRO/SMO$ multilayer series, we have measured
the value of $\psi $ from the $103$ asymmetric reflection. The values of $%
\sin ^2\psi _{103}$ for different bilayer thickness as a function of ($\frac %
1N$) is shown in the Fig. 5(c). The plot shows excellent agreement to eq.
6b. Using $a_{SRO}$ and $a_{SMO}$ as the bulk value and the value of $d_o$
calculated from Fig. 5(a), we have calculated the values of strain
components. The values of $\varepsilon _{11}$ and $\varepsilon _{33}$ are $%
4.0876$ and $-$ $0.678$ respectively. To compare these strain components, we
have plotted $d_{103}$($\phi \psi $) with the corresponding $\sin ^2\psi
_{103}$ for various samples in the inset of Fig. 5(c). The $d_{103}$($\phi
\psi $) and $\sin ^2\psi _{103}$ of these series satisfies eq. (4), and the
values of $\varepsilon _{11}$ and $\varepsilon _{33}$ are $4.419$ and $-$ $%
0.69$, respectively. These values are consistent with the values of strain
components calculated from the Fig 5(c). Using the values of $d_o$($%
\varepsilon _{11}$ $-$ $\varepsilon _{33}$) and $d_o$ ($\varepsilon _{33}$ $%
+ $ $1$) obtained from the inset of Fig. 5(c) and the slope and intercept of
Fig 5(c) in eq. 6b, we have calculated the value of strained lattice
parameter of $a_{SRO}$ and $a_{SMO}$ along the $00l$ direction. The values
of $a_{SRO}$ and $a_{SMO}$ are $3.99$ $\AA $ and $3.864$ $\AA $,
respectively, confirming the expansion and the compression in the
out-of-plane direction.

In $SRO/SMO$ multilayer structure, the out-of-plane direction has alternate
stacking of $RuO_6$ and $MnO_6$. In a superlattice with $n$ $=$ $1$ the
out-of-plane lattice parameter is $4.003$ $\AA $ which is larger than the
lattice parameter of the constituents as well as the substrate. This state
of strain indicates the elongation of these octahedra along the $c$-axis.
This superlattice has a larger strain state compare to the single layer of $%
SRO$ film, although the total thickness of the structure is larger than a
strain relaxed film ($150$ \AA\ \cite{25}). The lower strain-relaxed
thickness of $SRO$ on $STO$\ and larger difference in the lattice parameter
between the two compounds of the superlattice suggest that the modulation of
bilayer strain is the larger contribution to the total strain in the
superlattice. As the bilayer thickness, i.e. the spacer layer thickness
increases, the strain level in the bilayer relaxed and the distortion of
these octahedra decreases. This strain at room temperature due to the
interfaces is analyzed by $\sin ^2\psi $ methods. The strain in eq. (4)
depends on the $hkl$ orientations provided the strain is biaxial and
uniform. However, eq. (4) is valid for a thin film with single interfaces
and is not restricted to whether the strain is due to the volume conserving
modification or not. Thin films of transition metal compounds have been
observed to have strain gradient along the growth direction. The presence of
small step and terraces on the surface of the substrate may also induce
non-uniformity on the in-plane strain. This suggest that the value of $\sin
^2\psi $ may not follow eq. (4) for arbitrary $hkl$ orientations. For this
reason, the samples are studied along $10l$ orientations. The values of
strain components are similar to that of the $1000$ $\AA $\ thick film of $%
SRO$ on $STO$ seen in Fig. 5(b). In eq. (4) the parameters that include the
stacking nature of the samples are $d_0$, $E$ and $\nu $. In the strain
calculation, we have used the same $d_0$ for all samples though the average
bilayer lattice parameter is different. This could be the reason that we
could not extract any signature of the strain gradient along the $10l$
direction using eq. (4). Thus, we consider only the $103$ direction and
compared $d_{103}$($\phi \psi $) with $\sin ^2\psi _{103}$ of the samples
with different spacer layer thickness (Fig.5c). The linearly dependent of $%
d_{103}$($\phi \psi $) with $\sin ^2\psi _{103}$ for different sample allows
us to calculate the strain. The values of strain components are two times
larger than the values calculated along $10l$ direction. Also we have
calculated the values of strain components from $\sin ^2\psi _{103}$ using
eq. 6b. Both the calculations show approximately same values of strain. The
sign of strain components in the multilayer is similar to that of the strain
components of the $SRO$ thin film. This suggests that the in-plane tensile
strain induced on $SMO$ due to $20$ u.c. thick $SRO$ is not so strong to
overcome the substrate-induced strain. The strained out-of-plane lattice
parameter of $SRO$ and $SMO$ calculated from eq. (6) indicates the volume
conserving distortion of $SRO$, whereas the distortion in $SMO$ does not
conserve its volume even if it retains its cubic symmetry. At the
interfaces, the modified structure of $SMO$ is stabilized in the pseudocubic
phase, and suppress the strength of the in-plane tensile strain. In the
multilayer the interfaces between the constituents modulate the
substrate-induced strain which keeps the strain coherency in the sample. As
the bilayer thickness increases, the substrate-induced strain relaxes and it
is reflected in the out-of-plane lattice parameter of the multilayer.

The zero-field temperature-dependent resistivity ($\rho $) of these
superlattices are shown in the Fig. 6. The resistivity of $1000$ $\AA $\
thick film of $SRO$ is metal-like in the entire temperature range with
resistivity anomaly at $150$ $K$\cite{18}. While the resistivity of the
superlattice with $1$ $u.c.$ thick $SMO$ layer below room temperature is
metal-like with a resistivity minima at $20$ $K$ and below $20$ $K$ the
resistivity is insulator like. As the $SMO$ layer thickness increases the
resistivity minima shifted towards the higher temperature and $\rho $($T$)
shows an insulator-to-metal transition. This indicates the presence of
interface effect due to the $3D$ coordination of $Ru$ and $Mn$ ions, in the $%
\rho $($T$), though the top layer is a $20$ $u.c.$ thick $SRO$. The
resistivities in the inset of Fig. 6 at $10$ $K$ and $300$ $K$ of these
superlattices show continuous increase in its magnitude with the increase of 
$SMO$ layer thickness. For the sample with lower $SMO$ layer thickness where
the strain is larger (Fig. 4b), the change in the magnitude of the
resistivity is negligible. Although the transport measurement contains the
information of the interfaces, the effect of strain is dominated by the
magnetic state of the mobile carrier and the insulating nature of $SMO$
layer.

In conclusion, we have grown superlattices consisting of $20$ $u.c.$ thick $%
SrRuO_3$ and $n$ $u.c.$ $SrMnO_3$ where $n$ varies from $1$ to $20$ grown on
($001$)-oriented $SrTiO_3$ utilizing the pulsed laser deposition technique.
The evolution of the lattice parameters, the crystallinity and the epitaxy
of the films are evaluated as a function of the number of $SrMnO_3$ unit
cells using $X$-ray diffraction and transmission electron microscopy. We
have also studied the state of strain at the interfaces and the structural
coherency using the $\sin ^2\psi $ method. The superlattices show larger
residual strain compared to the single layer film of $SRO$ suggesting that
the lattice stiffening from interfacial strain and inhibiting dislocation by
composition modulation. These results bring new insights on the interfacial
stress measurements of oxide multilayers that can be used to control the
physical properties at the level of the atomic scale.

\bigskip

Acknowledgments:

We thank M.\ Morin for the preparation of samples for TEM\ cross-section and
J.\ Lecourt in the targets synthesis. We also thank Dr.\ H.\ Eng for careful
reading of the article.

We greatly acknowledged financial support of Centre Franco-Indien pour la
Promotion de la Recherche Avancee/Indo-French Centre for the Promotion of
Advance Research (CEFIPRA/IFCPAR) under Project N${{}^{\circ }}$%
2808-1.

\newpage Figures captions:

\bigskip

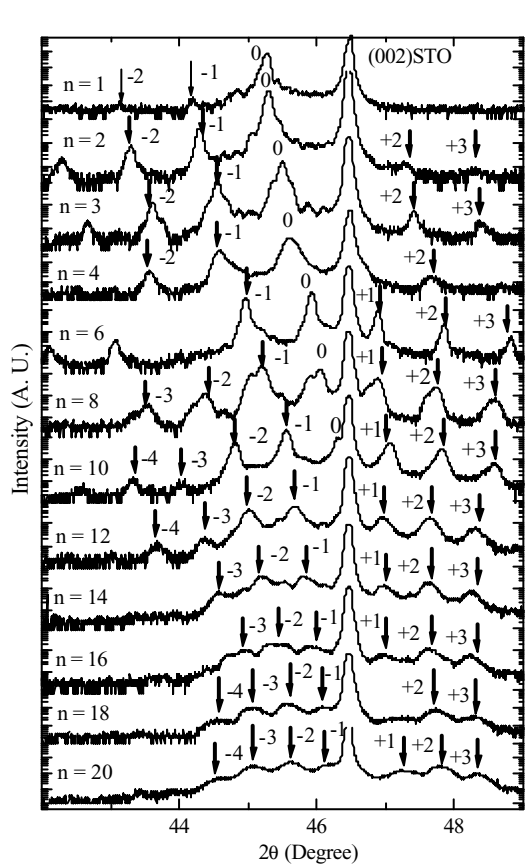
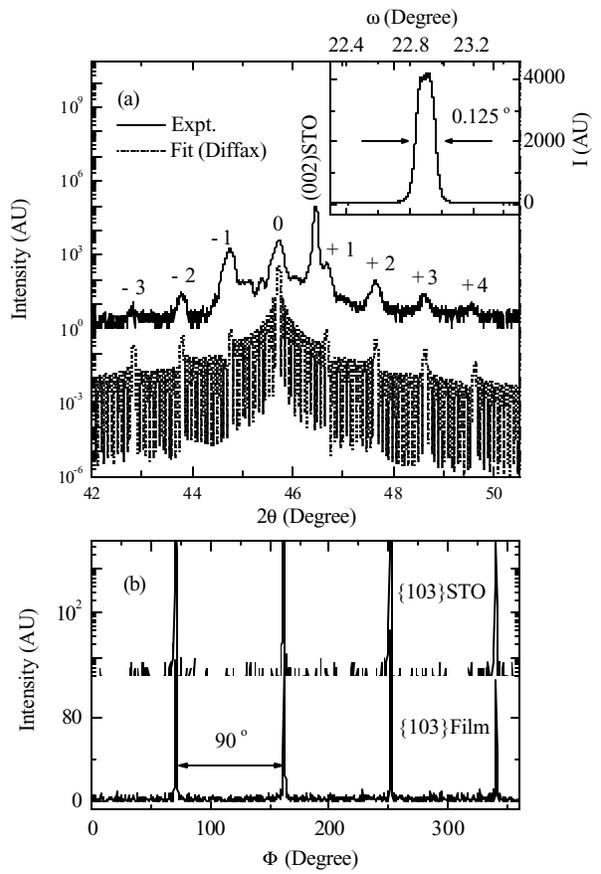
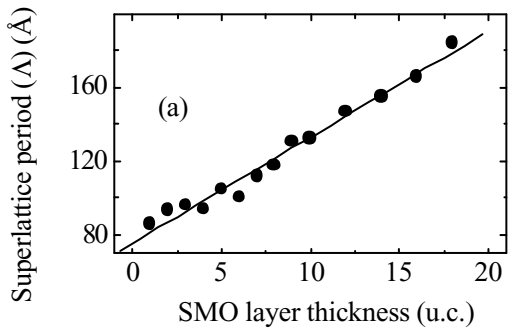
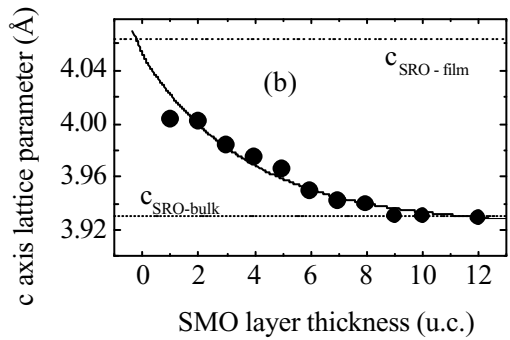
Fig.1: $\Theta $-2$\Theta $ scan recorded around the $002$ reflection of $%
STO $ for various multilayer ($SRO$)$_{20}$($SMO$)$_n$ ($n=1-20$). Note the
presence of satellites peaks (denoted by arrows) of several orders (from -3
to +3) around the main (fundamental) peak (order 0) attesting to the
formation of superlattices.

Fig.2(a): $\Theta $-2$\Theta $ scan around the $002$ reflection of $STO$ for
a multilayer ($SRO$)$_{20}$($SMO$)$_5$. The calculated intensity using
Diffax program is also indicated.\ Note the perfect agreement between
experimental and calculated intensity. The inset depicts the $\omega $-scan
recorded around the main peak for the same film.\ The low value of the $FWHM$%
\ close to $0.12{{}^{\circ }}$ confirms the high quality of the superlattice.

Fig.2(b): $\Phi $-scans recorded around the \{$103$\} of the film and the
STO\ substrate showing a 4-fold symmetry and an in-plane alignment.

Fig.3(a) Electron diffraction of a cross-section for a ($SRO$)$_{20}$($SMO$)$%
_5$ multilayer taken along the [010] direction.\ The inset is the
enlargement of the $001$ spot showing one satellite spot (SL) resulting from
the superstructure.

Fig.3(b): Overall cross-section image showing the $STO$\ substrate and the
superlattice ($SRO$)$_{20}$($SMO$)$_5$.\ The inset is an enlargement showing
the stacking.\ The $SRO$\ and $SMO$\ layers are clearly visible.\ The arrows
indicate the substrate-film interface.

Fig.4(a): Evolution of the superlattice period ($\Lambda $) as a function of
the number of $SMO$ layer calculated from the position of the satellite
peaks of Fig.1 (see text for details). The solid line is the fit to the data.

Fig.4(b): Evolution of the average $c$-axis lattice parameter ($=\Lambda
/(20+n)$) as a function of the number of $SMO$ layers, calculated from the
position of the satellite peaks of Fig.1 (see text for details). The line is
only a guide for the eyes. The $c$-axis value of the bulk $SRO$ as well as
the $c$-axis value obtained for a $20$ $u.c.$ thick $SRO$\ film are also
indicated for comparison.

Fig.5(a): $d_{10l}$ vs. sin$^2\psi _{10l}$ ($l=1,2,3$ and $4$) for ($SRO$)$%
_{20}$($SMO$)$_n$ with $n=1$ and $n=12$. The solid line is the fit to the
data.

Fig.5(b): Evolution of the in-plain strain ($\varepsilon _{11}$) and
out-of-plane strain ($\varepsilon _{33}$) as a function of the inverse of
the bilayer unit cell ($1/N$). The line is only a guide for the eyes.

Fig.5(c): Evolution of the inverse of the bilayer unit cell ($1/N$) as a
function of sin$^2\psi _{103}$. The inset depicts the evolution of $d_{103}$
vs. sin$^2\psi _{103}$ for different multilayers. The solid lines in the
figure are the fit to the data.

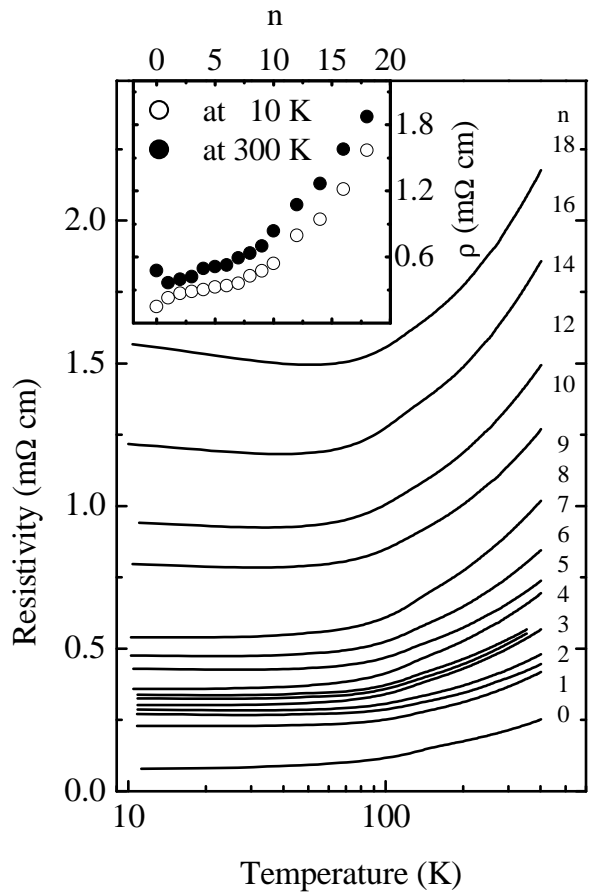
Fig.6: Temperature dependent zero-field resistivity for different
multilayers. Inset shows the values of zero-field resistivities at $10$ $K$
and $300$ $K$ of these multilayers.

\end{document}